\documentclass[aps,prx,twocolumn,superscriptaddress,longbibliography]{revtex4-1}

\usepackage{amsmath}
\usepackage{amssymb}
\usepackage{graphicx}
\usepackage[dvipsnames]{xcolor}
\usepackage[colorlinks,linkcolor=blue,citecolor=blue,urlcolor=blue]{hyperref}
\usepackage{stix}
\usepackage{tikz}

\definecolor{lime}{HTML}{A6CE39}
\DeclareRobustCommand{\orcidicon}{
  \begin{tikzpicture}
  \draw[lime, fill=lime] (0,0) 
  circle [radius=0.16] 
  node[white] {{\fontfamily{qag}\selectfont \tiny ID}};
  \draw[white, fill=white] (-0.0625,0.095) 
  circle [radius=0.007];
  \end{tikzpicture}
  \hspace{-2mm}
}

\foreach \x in {A, ..., Z}{\expandafter\xdef\csname orcid\x\endcsname{\noexpand\href{https://orcid.org/\csname orcidauthor\x\endcsname}
      {\noexpand\orcidicon}}
}

\begin{document}

\renewcommand{\vec}[1]{\mathbf{#1}}
\newcommand{\iu}{\mathrm{i}}
\newcommand{\hc}{\hat{c}}
\newcommand{\hcd}{\hat{c}^\dagger}
\newcommand{\en}{\varepsilon}
\newcommand{\gvec}[1]{\boldsymbol{#1}}
\renewcommand{\pl}{\parallel}

\newcommand{\new}[1]{\textcolor{WildStrawberry}{#1}}

\author{Michael Sch\"uler\orcidA{}}
\email{michael.schueler@psi.ch}
\affiliation{Condensed Matter Theory Group, Paul Scherrer Institute, CH-5232 Villigen PSI, Switzerland}
\affiliation{Laboratory for Materials Simulations, Paul Scherrer Institute, CH-5232 Villigen PSI, Switzerland}
\affiliation{Department of Physics, University of Fribourg, 1700 Fribourg, Switzerland}
\author{Thorsten Schmitt}
\affiliation{Photon Science Division, Paul Scherrer Institute, CH-5232 Villigen PSI, Switzerland}
\author{Philipp Werner\orcidC{}}
\affiliation{Department of Physics, University of Fribourg, 1700 Fribourg, Switzerland}

% Include your paper's title here

\title{Probing magnetic orbitals and Berry curvature with circular dichroism in resonant inelastic X-ray scattering} 

\begin{abstract}
Resonant inelastic X-ray scattering (RIXS) 
can probe localized excitations at selected atoms in materials, including particle-hole transitions between the valence and conduction bands. These transitions are governed by fundamental properties of the corresponding Bloch wave-functions, including orbital and magnetic degrees of freedom, and quantum geometric properties such as the Berry curvature. In particular, orbital angular momentum (OAM), which is closely linked to the Berry curvature, can exhibit a nontrivial momentum dependence. We demonstrate how information on such OAM textures can be extracted from the circular dichroism in RIXS. Based on accurate modeling with first-principles treatment of the key ingredient -- the light-matter interaction -- we simulate dichroic RIXS spectra for the prototypical transition metal dichalcogenide MoSe$_2$ and the two-dimensional topological insulator 1T$^\prime$-MoS$_2$. Guided by an intuitive picture for the optical selection rules, we discuss how the momentum-dependent OAM manifests itself in the dichroic RIXS signal if one controls the momentum transfer. Our calculations are performed for typical experimental geometries and parameter regimes, 
and demonstrate the possibility of observing the predicted circular dichroism in forthcoming experiments. Thus, our work establishes a new avenue to observing Berry curvature and topological states in quantum materials.
\end{abstract}

\maketitle 

\section*{Introduction}

The microscopic quantum geometry of Bloch electrons is one of the most remarkable properties of quantum materials, giving rise to fascinating macrosopic effects. Notatable examples are topological insulators and superconductors~\cite{hasan_colloquium:_2010,qi_topological_2008-2}, where the integral over the Berry curvature yields an integer -- the Chern number -- that is fundamentally connected to edge models and transport properties. 
More generally, the momentum-resolved Berry curvature plays a fundamental role for optical properties of solids, including intriguing effects in high-harmonic generation~\cite{li_probing_2013,tancogne-dejean_atomic-like_2018}, the valley Hall effect~\cite{barre_spatial_2019} and nonlinear Hall effect~\cite{xu_electrically_2018-1,ma_observation_2019-1}. 

Measuring the Berry curvature texture has proven difficult. The most promising route up to now is angle-resolved photoemission spectroscopy (ARPES)~\cite{damascelli_probing_2004,lv_angle-resolved_2019} with circularly polarized photons. %In particular, 
For example, 
the circular dichroism (CD) provides insights into the chirality in graphene~\cite{gierz_graphene_2012} and topological insulators~\cite{wang_observation_2011} and into the Berry curvature in two-dimensional materials~\cite{razzoli_selective_2017,cho_experimental_2018,cho_studying_2021, schuler_local_2020-1} and in Weyl semimetals~\cite{unzelmann_momentum-space_2021}. Circular photons couple to the texture of the orbital angular momentum (OAM) of the Bloch states~\cite{souza_dichroic_2008-1,resta_electrical_2010}, which, in turn, is closely connected to the Berry curvature~\cite{xiao_berry_2010,schuler_local_2020-1}, although  the precise relation can be complicated~\cite{ma_chiral_2015,unzelmann_momentum-space_2021}. 
The connection between the OAM texture and the measured CD is governed by the photoemission matrix elements, which depend on the details of the final states~\cite{bentmann_strong_2017,beaulieu_revealing_2020}, the specifics of the experimental geometry~\cite{schonhense_circular_1990}, 
mixing of different orbital channels~\cite{moser_huygens_2022}, and the short escape depth of the photoelectrons~\cite{moser_experimentalists_2017}. Especially for bulk materials, the link between the CD signal and OAM is obscured by these complications, rendering the extraction of Berry curvature a difficult task.

One of the most powerful techniques for revealing the magnetic properties of materials -- including OAM -- is resonant inelastic X-ray scattering (RIXS)~\cite{ament_resonant_2011}. This technique provides access to excitation in various degrees of freedom, including magnetic excitations such as magnons, as well as orbital excitations~\cite{schlappa_spinorbital_2012}. Controlling the polarization of the incoming X-ray photons allows to select or enhance certain excitations as governed by the selection rules. For $d$ electron systems, this selectivity
has been exploited to maximize spin-flip excitations~~\cite{van_veenendaal_polarization_2006} to map out the magnetic degrees of freedom~\cite{ament_theoretical_2009}. The quantum geometry of the Bloch states can impose further restrictions on the the final states~\cite{ahn_riemannian_2022}. Based on this idea Ref.~\cite{kourtis_bulk_2016} has demonstrated how the chirality in Weyl semimetals leads to linear dichroism in RIXS. 
Thanks to recent advances in measuring RIXS with circularly polarized photons, it becomes possible to reveal OAM and thus measure local magnetic orbital exciations. Transitions between magnetic orbitals are a direct fingerprint of
OAM localized at the targeted atoms. Hence, CD in RIXS (CD-RIXS) provides an unprecedented probe for the chirality of the electronic wave-function and the breaking of time-reversal symmetry. CD-RIXS is conceptionally similar to X-ray magnetic CD (XMCD) in X-ray absorption, and it has be used to study magnetic materials~\cite{zimmermann_1s2p_2018,kotani_resonant_2005,magnuson_large_2006,miyawaki_dzyaloshinskii-moriya_2017,iwazumi_magnetic_2003}, spin-flip processes~\cite{elnaggar_magnetic_2019}, or anti-ferromagnetic orbital order~\cite{marra_unraveling_2012}. However, the link between the local OAM \emph{texture} and CD-RIXS has not been explored.

\begin{figure*}[t]
\centering\includegraphics[width=\textwidth]{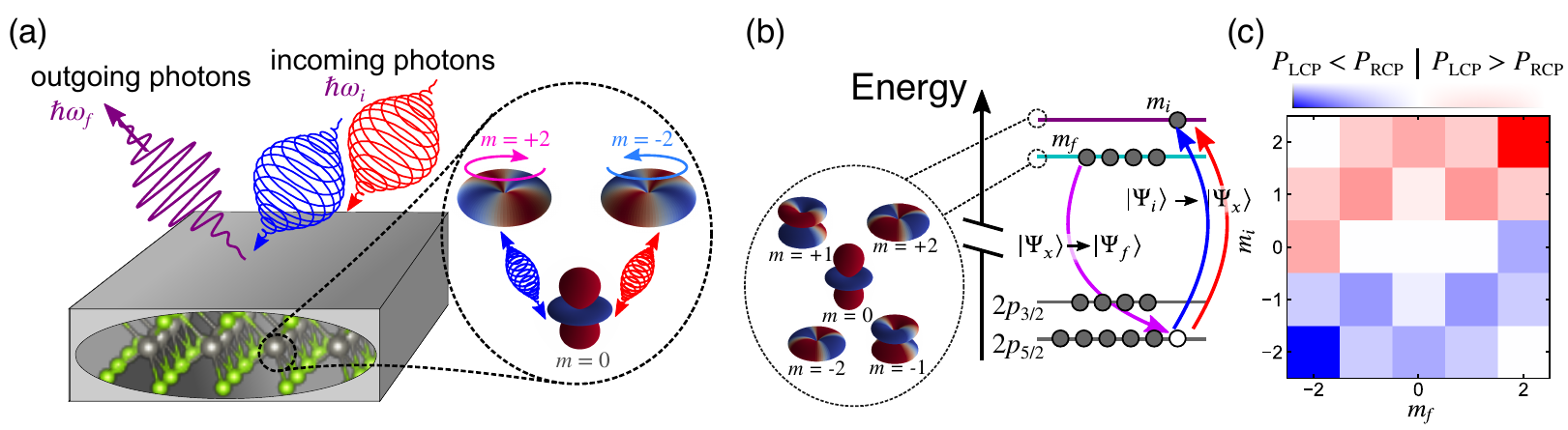}
\caption{\textbf{Sketch of the setup and dichroic selection selection rules.} (a) We investigate resonant inelastic X-ray scattering (RIXS) with circularly polarized photons (with energy $\hbar \omega_i$) impinging on a sample, while scattered photons (with energy $\hbar \omega_f$) are detected without resolving their polarization. This setup is sensitive to local magnetic orbital excitations which are controlled by the circular polarization of the incoming light. 
(b) Sketch of $d$-$d$ excitations in the RIXS process: upon absorbing a circularly polarized photon, an electron is promoted from the $2p$ core levels
to an unoccupied state in the $d$ manifold with magnetic quantum number $m_i$. A filled $d$ orbital (quantum number $m_f$) relaxes and fills the core hole. Effectively, this corresponds to an excitation within the $d$ orbital space. (c) Circular dichroism of the $d$-$d$ transition probability $P$ resolved with respect to the magnetic quantum number of the $d$ orbitals.
\label{fig:sketch}}
\end{figure*}

Because Berry curvature is typically tied to local OAM -- especially in transition-metal dichalcogenides (TMCDs), which are a prominent class of two-dimensional materials~\cite{novoselov_two-dimensional_2005,novoselov_2d_2016,kou_two-dimensional_2017} -- CD-RIXS can reveal quantum geometric information beyond basic magnetic properties. 
This novel aspect is the focus of this work. We show how CD-RIXS provides a fingerprint of magnetic orbital excitations in paradigmatic TMDCs. Exploiting the momentum resolution of RIXS, we explore how the \emph{momentum-dependent} OAM texture manifests itself in the CD-RIXS signal, using accurate modeling with first-principle input. We focus on monolayer MoSe$_2$ to demonstrate how the valley-dependent OAM can be mapped out, and on 1T$^\prime$-MoS$_2$, %currently one of the most promising 
which is a representative of the class of intensively studied 
two-dimensional topological insulators~\cite{tang_quantum_2017,fei_edge_2017,wu_observation_2018,marrazzo_relative_2019}, in which Berry curvature and an associated dipole can be induced and tuned by applying an out-of-plane electric field~\cite{xu_electrically_2018-1}. We discuss how this intriguing scenario can be observed with CD-RIXS, underlining the power of CD-RIXS as a novel probe for quantum materials.

\section*{Results}

In the RIXS process, a core electron is promoted to the unoccupied bands upon absorbing an X-ray photon with energy $\hbar\omega_i$. The core hole is filled by a relaxing valence electron which emits an outgoing photon with lower energy $\hbar \omega_f$, such that the system absorbs the energy $\hbar\Delta\omega = \hbar\omega_i - \hbar\omega_f$ (see illustration in Fig.~\ref{fig:sketch}). The cross section of this process is described theoretically by the Kramers-Heisenberg formula
\begin{align}
  \label{eq:kramers_heisenberg_1}
  I(\omega_i,\vec{q}_i,\omega_f,\vec{q}_f) = \sum_f \left|A_{fi}(\vec{\omega}_i,\vec{q}_i,\vec{q}_f)\right|^2
  \delta(E_i + \Delta \omega - E_f) \ .
\end{align} 
We use atomic units (a.u.) here and in what follows.
The incoming (outgoing) X-ray photon carries the momentum $\vec{q}_i$ ($\vec{q}_f$). Thus the total momentum $\vec{q} = \vec{q}_f - \vec{q}_i$ is transfered to the system. Controlling the momentum transfer $\vec{q}$ via the experimental geometry provides momentum-resolved information, and allows to map out the dispersion of fundamental quasi-particle excitations such as phonons, magnons and orbital excitations.

The RIXS signal~\eqref{eq:kramers_heisenberg_1} is determined by the RIXS amplitude
\begin{align}
  \label{eq:rixs_ampl_1}
  A_{fi}(\vec{\omega}_i,\vec{q}_i,\vec{q}_f) = \sum_x \frac{\langle \Psi_f | \hat{\Delta}^\dagger_f | \Psi_x\rangle
  \langle \Psi_x | \hat{\Delta}_i | \Psi_i\rangle}{E_x - E_i - \omega_i - i\Gamma}\ ,
\end{align}
where the $|\Psi_x\rangle$ denote all intermediate states with a single core hole (energy $E_x$), while $\hat{\Delta}_i = $ ($\hat{\Delta}_f$) represents the light-matter interaction involving the incoming (outgoing) photon. We go beyond the dipole approximation by incorporating the momentum of the photons and representing the dipole operator by the momentum operator $\hat{\vec{p}}$.
The short lifetime $\tau_\mathrm{ch}$ of the core hole enters Eq.~\eqref{eq:rixs_ampl_1} as the broadening parameter $\Gamma=1/(2\tau_\mathrm{ch})$.

Evaluating the RIXS cross section~\eqref{eq:kramers_heisenberg_1} from Eq.~\eqref{eq:rixs_ampl_1} is an intricate problem, as the interaction of the electrons both in the core levels and in the valence bands plays a role. Furthermore, the creation of a strongly localized core hole is accompanied by strong electrostatic interactions, which give rise to strongly bound excitons 
and can trigger collective excitations such as phonons and plasmons. In this work the light-matter interaction is the key ingredient, while details of the core electrons only play a minor role. Therefore, we treat the band electrons on the level of density functional theory (DFT), which captures the detailed electronic structure, while tabulated values for the deep core levels are used~\cite{clementi_atomic_1963,clementi_atomic_1967}. In practice, DFT can capture the multiplet structure of the intermedate and final states only approximetely. While the final states correspond to particle-hole excitations which can be described well within DFT, the relaxation effects in the presence of the core holes render the intermediates states more correlated. The combination with a linear-response treatment of relaxation effects has been shown to yield reasonable agreement with experiments~\cite{hanson-heine_kohn-sham_2017,fouda_simulation_2018} for molecules. For weakly and moderately correlated systems such as perovskites~\cite{zhuo_spectroscopic_2018}, solving the Bethe-Salpeter equation (BSE) to describe the intermediates states has been successful~\cite{vinson_bethe-salpeter_2011}. Here we use a simplified approach in a similar spirit, where core-hole excitons are included in our theory by considering the core-hole Coulomb interaction $U^c$. Employing the mean-field approximation allows to explictly solve for the intermediates states $|\Psi_x\rangle$ using the exciton formalism~\cite{sangalli_ab-initio_2018}, which are then used to evaluate the RIXS amplitude~\eqref{eq:rixs_ampl_1}.

By using the mean-field approximation the position of the absorption edge (determined by the exciton binding energy) is only described qualitatively. However, the essence of core-hole relaxation effects and their impact on the CD-RIXS signal are included in our theory. We have ascertained the robustness of our results by varying $U^c$. Together with the first-principle treatment of the light-matter action, our key findings are expected to hold when employing more refined simulations.

\subsection*{Orbital excitations with circular photons}

% \begin{figure}[t]
% \centering\includegraphics[width=\columnwidth]{figure_dorbitals_rtensor.pdf}
% \caption{\textbf{RIXS process and selection rules.} (a) Sketch of $d$-$d$ excitations in the RIXS process: upon absorbing a circularly polarized photon, an electron is promoted from the $2p$ core levels
% to an unoccupied state in the $d$ manifold with magnetic quantum number $m_i$. A filled $d$ orbital (quantum number $m_f$) relaxes and fills the core hole. Effectively, this corresponds to an excitation within the $d$ orbital space. (b) Circular dichroism of the $d$-$d$ transition probability $P$ resolved with respect to the magnetic quantum number of the $d$ orbitals.
% }
% \label{fig:illustration_dorb}
% \end{figure}

Before presenting the simulated RIXS spectra for specific materials, we discuss how circularly polarized photons can reveal local magnetic orbital excitations. To this end we analyze the core ingredient of the RIXS amplitude~\eqref{eq:rixs_ampl_1}: the transition matrix elements. To induce transitions to magnetic $d$ orbitals, the core states  need to have $p$ orbital character (at least for the dominant dipole transitions). The spin-orbit coupling (SOC) in the core splits the core levels in the $2p$ shell into two distinct groups, $2p_{5/2}$ (coined L$_2$ edge) and $2p_{3/2}$ (L$_3$ edge). We focus on the L edge here. Exciting from the $3p$ shell (which splits into the M$_2$ and M$_3$ edge, respectively) is similar in principle, albeit enhanced correlations effects complicate the picture.
% Calculations were also performed for the $3p$ shell (which splits into he M$_2$ and M$_3$ edge, respectively), with no major differences in the CD signal except for an overall reduction of the intensity.

Assuming a fully local picture, we consider the transition (see Fig.~\ref{fig:sketch}(b)) from the ground state $|\Psi_i\rangle$ (no core holes, a single occupied $d$ orbital with magnetic quantum number $m_f$) to intermediate states $|\Psi_x\rangle$ (one core hole, additional electron in the $d$ shell with quantum number $m_i$) to final states $|\Psi_f\rangle$ ($d$ electron with quantum number $m_f$ filling the core hole). 
For illustrative purposes we %ignore the by taking 
use 
the ultra-short core-hole lifetime (UCL) approximation $\Gamma\rightarrow \infty$ here. We note, however, that the simulated RIXS spectra below are obtained from the full expression for the RIXS amplitude~\eqref{eq:rixs_ampl_1}.
Within the UCL,
the RIXS probability simplifies to
\begin{align}
  \label{eq:prob_simple}
  P(m_i,m_f) \propto \frac{1}{\Gamma^2}\left|\sum_x \langle \Psi_f | \hat{\Delta}^\dagger_f | \Psi_x\rangle
  \langle \Psi_x | \hat{\Delta}_i | \Psi_i\rangle\right|^2 .
\end{align}
The transition probability~\eqref{eq:prob_simple} depends on the polarization of the incoming (outgoing) light $\vec{e}_i$ ($\vec{e}_f$). As in experiments, we average over the polarization of the scattered photons. We can now understand which transitions $m_i\rightarrow m_f$ are driven by circularly polarized incoming photons. Figure~\ref{fig:sketch}(c) -- obtained within the independent-electron approximation in a geometry as below -- shows the difference $P_\mathrm{CD}(m_i,m_f) = P_\mathrm{LCP}(m_i,m_f) - P_\mathrm{RCP}(m_i,m_f)$ of the RIXS probability~\eqref{eq:prob_simple} with respect to left-hand circular polarization (LCP) and right-hand circular polarization (RCP). 
We observe the following general trends: (i) Except for the $m_i=0$ case, the circular dichroism is always positive (negative) for $m_i>0$ ($m_i < 0$). 
Hence, the dichroic signal $P_\mathrm{CD}(m_i,m_f)$ is a direct map of the local orbital angular momentum (OAM) of the \emph{unoccupied} states $m_i$. (ii) For $m_i=0$, one finds positive (negative) dichroism for $m_f=-2$ ($m_f=+2$). In this case the dichroic signal is sensitive to the magnetic state of the \emph{occupied} orbitals $m_f$. While the dichroic selection rules illustrated in Fig.~\ref{fig:sketch}(c) have been obtained for an noninteracting electron model, they are expected to also determine the chiral optical properties of the intermediate excitonic states, as has been shown for excitons in two-dimensional semiconductors~\cite{cao_unifying_2018,caruso_chirality_2021}.

\subsection*{Circular RIXS from monolayer MoSe$_2$}

\begin{figure*}[ht!]
\centering\includegraphics[width=\textwidth]{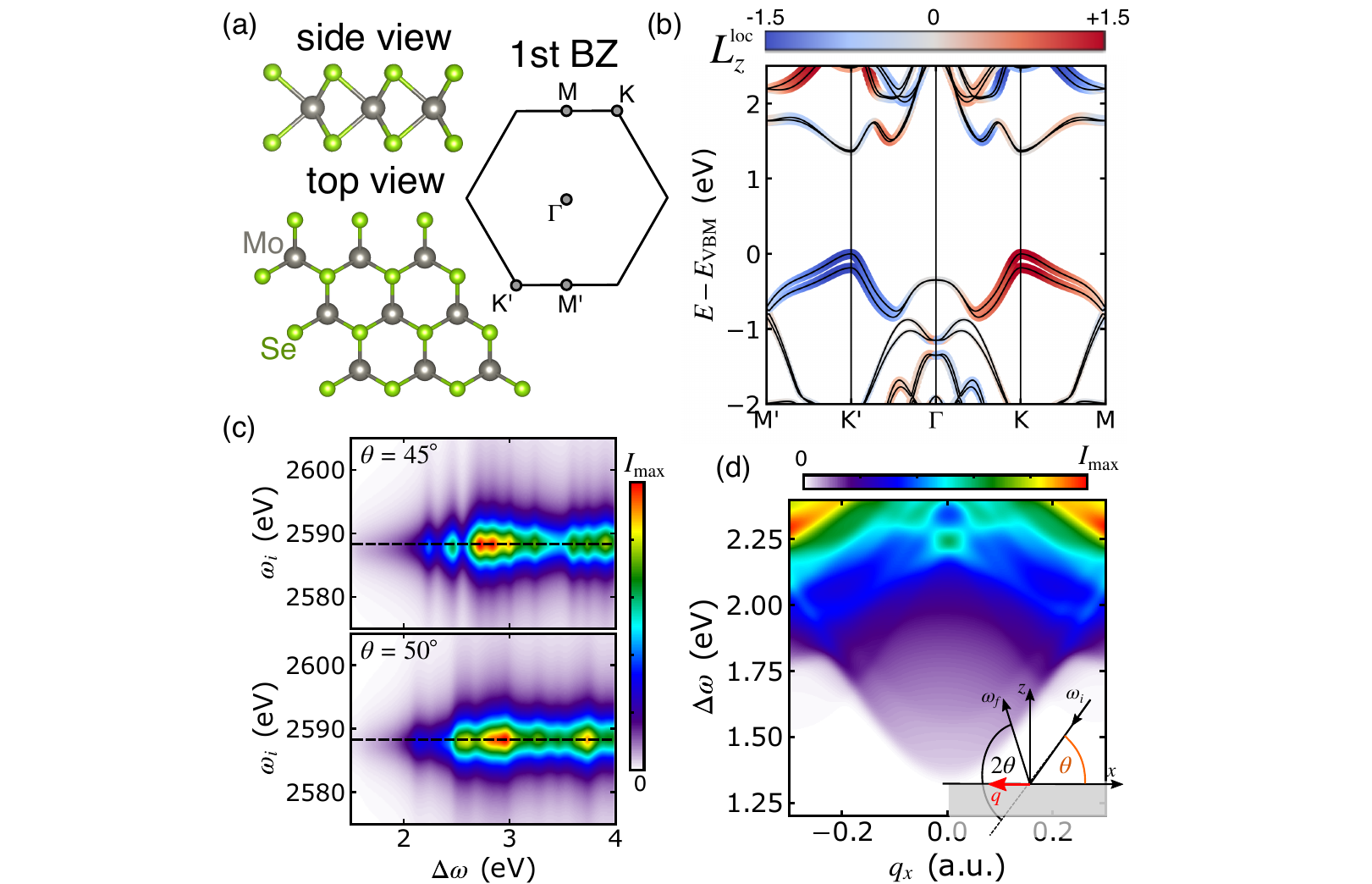}
\caption{\textbf{Properties and RIXS spectrum of monolayer MoSe$_2$.} (a) Sketch of the lattice structure of MoSe$_2$ and the first Brillouin zone including the high-symmetry points. (b) Band structure relative to the valence band maximum (VBM) along the indicated path. The colored fat bands represent the weight of the Mo-$d$ orbitals (size) and the local angular momentum (color map). (c) RIXS map showing the $d$--$d$ particle-hole excitations at incidence angle $\theta=45^\circ$ (top) and $\theta=50^\circ$ (bottom panel). (d) Dispersion of the particle-hole excitations at fixed incoming photon energy as indicated by the horizontal dashed line in (c). The inset on the bottom right illustrates the geometry of the calculation: the angle $2\theta=130^\circ$ is fixed, 
while varying $\theta$ corresponds to scanning over the in-plane momentum transfer $q$. The photon energy $\omega_i$ is tuned to the L$_2$ edge.
\label{fig:rixs_mose2}
}
\end{figure*}

With this intuition we can now investigate RIXS with circular photons from relevant $d$-electron materials. First we consider monolayer MoSe$_2$, a two-dimensional TMDC with remarkable spin polarization, Berry curvature, and OAM~\cite{fang_ab_2015}. The lattice structure and the first Brillouin zone (BZ) are sketched in Fig.~\ref{fig:rixs_mose2}(a). 
The band structure (see Fig.~\ref{fig:rixs_mose2}(b)) exhibits two top valence bands split by SOC with almost pure, opposite spin polarization at K and K$^\prime$, respectively. The electronic properties around the K and K$^\prime$ valleys -- the most important region due to the direct band gap -- is dominated by the Mo-$d$ orbitals, as shown by the width of the colored lines in Fig.~\ref{fig:rixs_mose2}(b). The orbital character of the top valence band around the K (K$^\prime$) point is dominated by the $d_{+2} = (d_{x^2-y^2} + i d_{xy})/\sqrt{2}$ ($d_{-2} = (d_{x^2-y^2} - i d_{xy})/\sqrt{2}$) orbital with an admixture of $d_{0}=d_{z^2}$ further away from the Dirac points. Hence, the OAM associated with the Mo atoms, $L^\mathrm{loc}_z$, is very pronounced (color coding in Fig.~\ref{fig:rixs_mose2}(b)). The bottom conduction bands are dominated by the $d_{z^2}$ orbital at K (K$^\prime$) with a growing contribution from $d_{+2}$ ($d_{-2}$) away from the valley center. The local OAM of the top valence band is the dominant contribution to the total OAM, which in turn determines the Berry curvature of the valence bands. Due to time-reversal symmetry, the Berry curvature and the OAM at K (K$^\prime$) has the opposite sign at K$^\prime$ (K).

We chose MoSe$_2$ as a representative of the TMDCs of type MX$_2$ because RIXS from the Mo $2p$ shell is straightforward to measure, as demonstrated in Ref.~\cite{thomas_resonant_2015} for various compounds. Furthermore, the photon energy in the tender X-ray regime allows for incidence angles close to normal incidence, which reduces geometric effects and enables an interpretation of the CD signal in terms of the magnetic quantum numbers $m_i$, $m_f$.  
Hence, with X-ray photons tuned to the L$_2$ (or L$_3$) edge of Mo, the RIXS signal from MoSe$_2$ is expected to qualitatively follow the scenario described by Fig.~\ref{fig:sketch}(c). To confirm this picture, we have calculated the RIXS spectra (L$_2$ edge) from Eq.~\eqref{eq:kramers_heisenberg_1} and Eq.~\eqref{eq:rixs_ampl_1} with first-principle input. Details are given in the Materials and Methods section. The only parameters are the local Coulomb interaction $U^c$ between the core hole and band electrons localized at the Mo sites and the inverse lifetime $\Gamma$. We fix $U^c=8$~eV and $\Gamma=3$~eV, which can be estimated from the effective charge $Z$ of the core states by assuming a quadratic scaling with $Z^{-2}$.
Figure~\ref{fig:rixs_mose2}(c) shows typical RIXS maps (polarization-avereged) for two different incidence angles $\theta$. 
The resonant behavior with respect to $\omega_i$ is due to the resonant structure in the RIXS amplitude~\eqref{eq:rixs_ampl_1}. Inspecting the dependence on the energy loss $\Delta \omega$ we notice a clear difference between the spectra for different $\theta$, which indicates a dispersion of the underlying excitations. To extract this dispersion of the particle-hole excitations in the Mo-$d$ orbitals, we focus on the resonant region (black dashed line in Fig.~\ref{fig:rixs_mose2}(c)) to maximize the intensity in analogy to experiments. The dispersion of the orbital particle-hole excitations is computed in analogy to typical experimental setups~\cite{schlappa_spinorbital_2012} (illustrated in the inset in Fig.~\ref{fig:rixs_mose2}(d)). The in-plane momentum transfer $q$ is determined by the incidence angle $\theta$, while the scattering angle $2\theta$ remains fixed. Varying $\theta$ and converting to $q$ then yields the dispersion presented in Fig.~\ref{fig:rixs_mose2}(d). In the absence of strong correlations in the valence bands, the dispersive RIXS signal in Fig.~\ref{fig:rixs_mose2}(d) can be understood as originating from transitions from the occupied valence bands to the conduction bands with momentum change $q_x$. In reality, exciton features would appear at $\Delta \omega$ below the band gap, which need to be separated out from the signal to focus on the particle-hole excitations, as discussed in Ref.~\cite{paris_strain_2020}.
For $q_x\approx 0$, the computed signal is dominated by almost vertical transitions across the direct band gap, while the combined dispersion of the top valence and bottom conduction band determines the dispersion of the orbital excitations.

\begin{figure*}[t]
\centering\includegraphics[width=\textwidth]{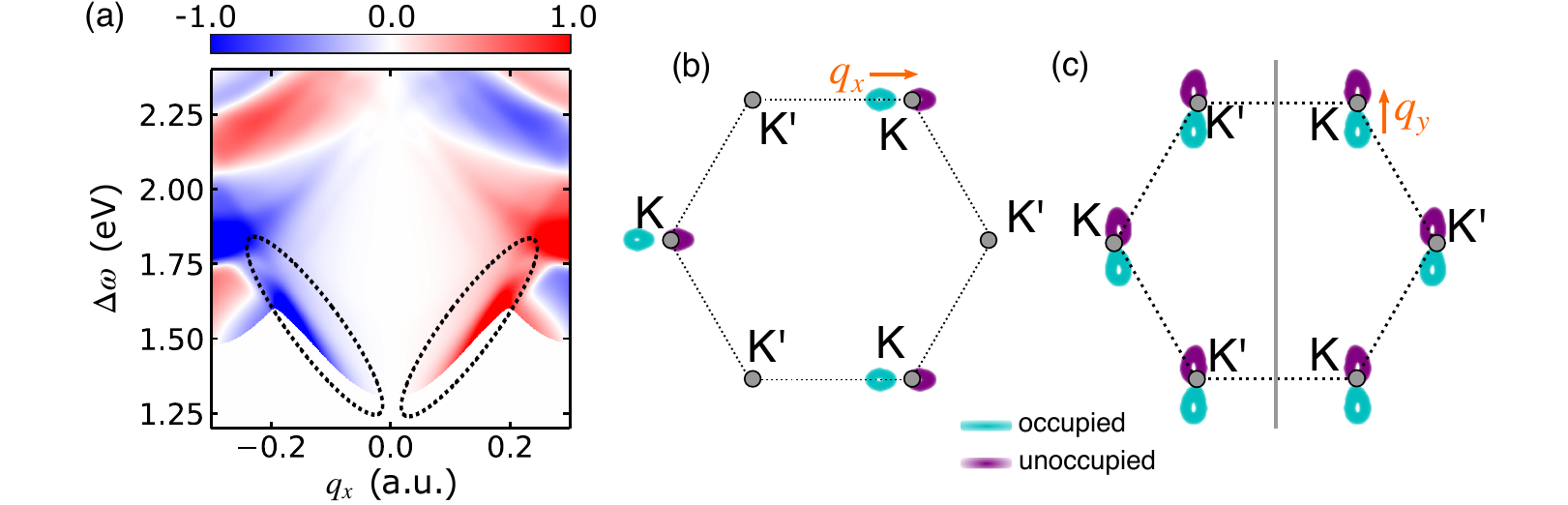}
\caption{\textbf{Circular dichroism in RIXS from MoSe$_2$.} (a) Normalized dircular dichroism of the RIXS cross section $(I_\mathrm{LCP} - I_\mathrm{RCP}) / (I_\mathrm{LCP} + I_\mathrm{RCP} )$ of monolayer MoSe$_2$. $I_\mathrm{LCP}$ ($I_\mathrm{RCP}$) denotes the RIXS intensity for left-ciruclar (right-circular) polarization. The incoming photon energy $\omega_i$ and the geometry is the same as in Fig.~\ref{fig:rixs_mose2}(d). The ellipses indicate the region where the OAM of the top valence band dominates the signal.
(b) Sketch of the phase space of the $d$--$d$ excitations permitted by energy and momentum conservation at fixed momentum transfer $q_x = 0.164$~a.u. and energy loss $\Delta \omega = 1.73$~eV. The scattering plane is spanned by the $x$ and the $z$ direction. (c) Analogous sketch for momentum transfer $q_y$ along the $y$ direction. The vertical grey line indicates the mirror plane.}
\label{fig:mose2_cdrixs}
\end{figure*}

At small momentum transfer the excitations are predominantly $d_{\pm 2} \rightarrow d_{z^2}$. From Fig.~\ref{fig:sketch}(c) we expect circular dichroism in the RIXS process for these orbital transitions, which --  due to the almost one-to-one correspondence between local OAM and Berry curvature -- reflects the Berry curvature of the top valence band. However, the total Berry curvature around K and K$^\prime$ is opposite, so these contributions cancel out when integrating over the BZ. In such a situation, can there still be circular dichroism in the RIXS process? The answer is affirmative, as demonstrated by Fig.~\ref{fig:mose2_cdrixs}(a). 

While the dichroic signal vanishes at $q_x=0$, there is pronounced dichroism for $q_x \ne 0$ that changes sign with $q_x$. This can be understood by inspecting the momentum and energy conservation of the RIXS process. Starting from a valence state with energy $\varepsilon_{\vec{k}v}$, excitations are only allowed to conduction states with energy $\varepsilon_{\vec{k}+\vec{q}c} = \varepsilon_{\vec{k}v} + \Delta \omega$. The breaking of inversion symmetry gives rise to anistropic electronic orbital textures with respect to $\vec{q}$. This anisotropy leads to an imbalance between the contributions from the inequivalent K and K$^\prime$ valleys, as illustrated in Fig.~\ref{fig:mose2_cdrixs}(b). In the case of the dichroic features highlighted by the dashed ellipses in Fig.~\ref{fig:mose2_cdrixs}(a), the RIXS signal is predominantly determined by transitions at the K$^\prime$ (K) points for $q_x > 0$ ($q_x < 0$), where the valence band exhibits $d_{-2}$ ($d_{+2}$) orbital character. Consistent with Fig.~\ref{fig:sketch}(c), positive circular dichroism is observed
for $q_x > 0$, and the result reverses upon reversing the sign of $q_x$. 

For larger $|q_x|$ other bands start contributing to the RIXS signal, and the simple picture of $d_{\pm 2} \rightarrow d_{z^2}$ transitions breaks down. Instead the circular dichroism is governed by the magnetic quantum number $m_i$ corresponding the OAM of the conduction band while the valence band character has almost no influence, consistent with the scenario (ii) outlined in the discussion of Fig.~\ref{fig:sketch}.

Changing the scattering plane to the $y$--$z$ plane on the other hand leads to exactly vanishing circular dichroism. As illustrated in Fig.~\ref{fig:mose2_cdrixs}(c), the contribution from the K and K$^\prime$ valleys is identical in this case. The reflection symmetry of the lattice structure with respect to the $y$ axis (Fig.~\ref{fig:rixs_mose2}(a)) manifests as a mirror symmetry in reciprocal space, as indicated by the grey vertical line in Fig.~\ref{fig:mose2_cdrixs}(c). As the mirror operation swaps K $\leftrightarrow$ K$^\prime$, there can be no difference of the contribution from K and K$^\prime$. 

Hence, by chosing the momentum transfer along the $x$ direction, one can achieve valley selectivity and extract the valley-resolved Berry curvature. Controlling the phase space of the particle-hole excitations by varying $\vec{q}$ is a universal principle that applies to systems with broken inversion symmetry.

\subsection*{Tracing the tunable Berry curvature dipole in 1T$^\prime$-MoS$_2$}

With the RIXS signal from the prototypical TMDC MoSe$_2$ qualitatively understood, we now investigate monolayer 1T$^\prime$-MoS$_2$, which is a quantum-spin Hall effect insulator (QSHI)~\cite{qian_quantum_2014}. The most studied representative of this class of TMDCs is 1T$^\prime$-WTe$_2$~\cite{qian_quantum_2014,wu_observation_2018,tang_quantum_2017,tang_quantum_2017}, which is currently considered one of the most robust monolayer QSHI systems~\cite{mounet_two-dimensional_2018,marrazzo_relative_2019}. However, its band gap is very sensitive to strain. 1T$^\prime$-MoS$_2$ on the other hand has a larger band gap, but is slightly less mechanically stable, which is why most experiments on the 1T$^\prime$ TMDCs have been performed on WTe$_2$. Since measuring RIXS from the W $2p$ core shell is challenging, we present results for the simulated RIXS spectra for 1T$^\prime$-MoS$_2$ here. The synthesis of monolayer 1T$^\prime$-MoS$_2$ has already been achieved~\cite{liu_phase-selective_2018}, so that measuring the RIXS signal from this material should be feasible. For completeness we also present results for 1T$^\prime$-WTe$_2$ in the Supplementary Materials~\cite{supplement}.

\begin{figure*}[t]
\centering\includegraphics[width=\textwidth]{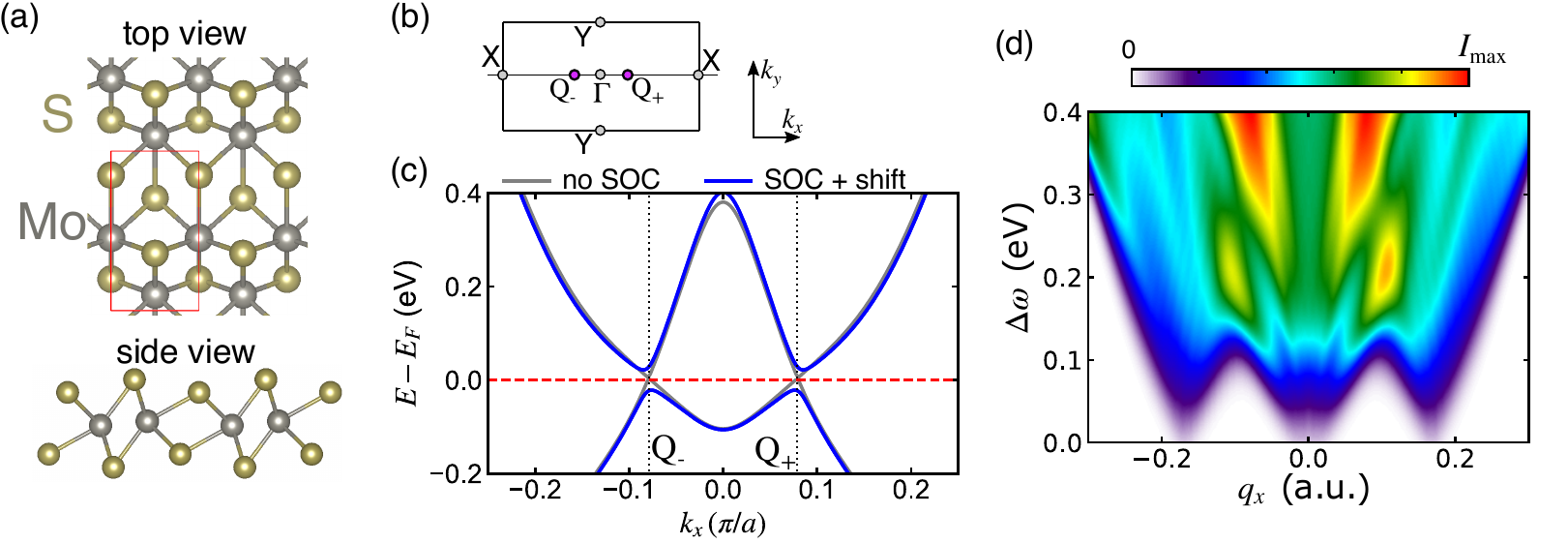}
\caption{\textbf{Electronic structure and RIXS spectrum of 1T$^\prime$-MoS$_2$.} (a) Sketch of the lattice structure. The red rectangle indicates the unit cell. (b) First BZ with important momenta. (c) Band structure with and without including SOC. Without SOC, there is a pair of Dirac cones at Q$_{\pm}$. (d) Simulated polarization-averaged RIXS intensity at temperature $T=20$~K. The scattering angle is fixed at $2\theta=130^\circ$.}
\label{fig:tpmos2}
\end{figure*}

The crystal structure (Fig.~\ref{fig:tpmos2}(a)) possesses inversion symmetry, which excludes any momentum-resolved OAM. The band structure along the $k_x$ direction (Fig.~\ref{fig:tpmos2}(b)) without SOC features a pair of Dirac cones located at the $Q_{\pm}$ points that gap out upon switching on SOC (see Fig.~\ref{fig:tpmos2}(c)). In contrast to other 1T$^\prime$ TMDCs, the electronic structure, including a good estimate of the band gap, can be obtained from DFT without the need for corrections~\cite{qian_quantum_2014}.
We computed the RIXS signal at the L$_2$ and L$_3$ edge in an analogous fashion as described for MoSe$_2$.
The calculated spectrum (Fig.~\ref{fig:tpmos2}(d)) for momentum transfer along the $x$ direction is a direct manifestion of the  specifics of the band structure: for small energy loss $\Delta \omega < 0.1$~eV there are vertical transitions ($q_x = 0$) and transitions at $q_x=\pm 0.18$~a.u., which correspond to excitations of the electrons around $Q_-$ ($Q_+$) to $Q_+$ ($Q_-$).
At low enough temperature (we set $T=20$~K) and with the energy resolution used in the simulations ($\sim 50$~meV), transitions below $\Delta \omega \approx 50$~meV are suppressed.

\begin{figure*}[ht!]
\centering\includegraphics[width=\textwidth]{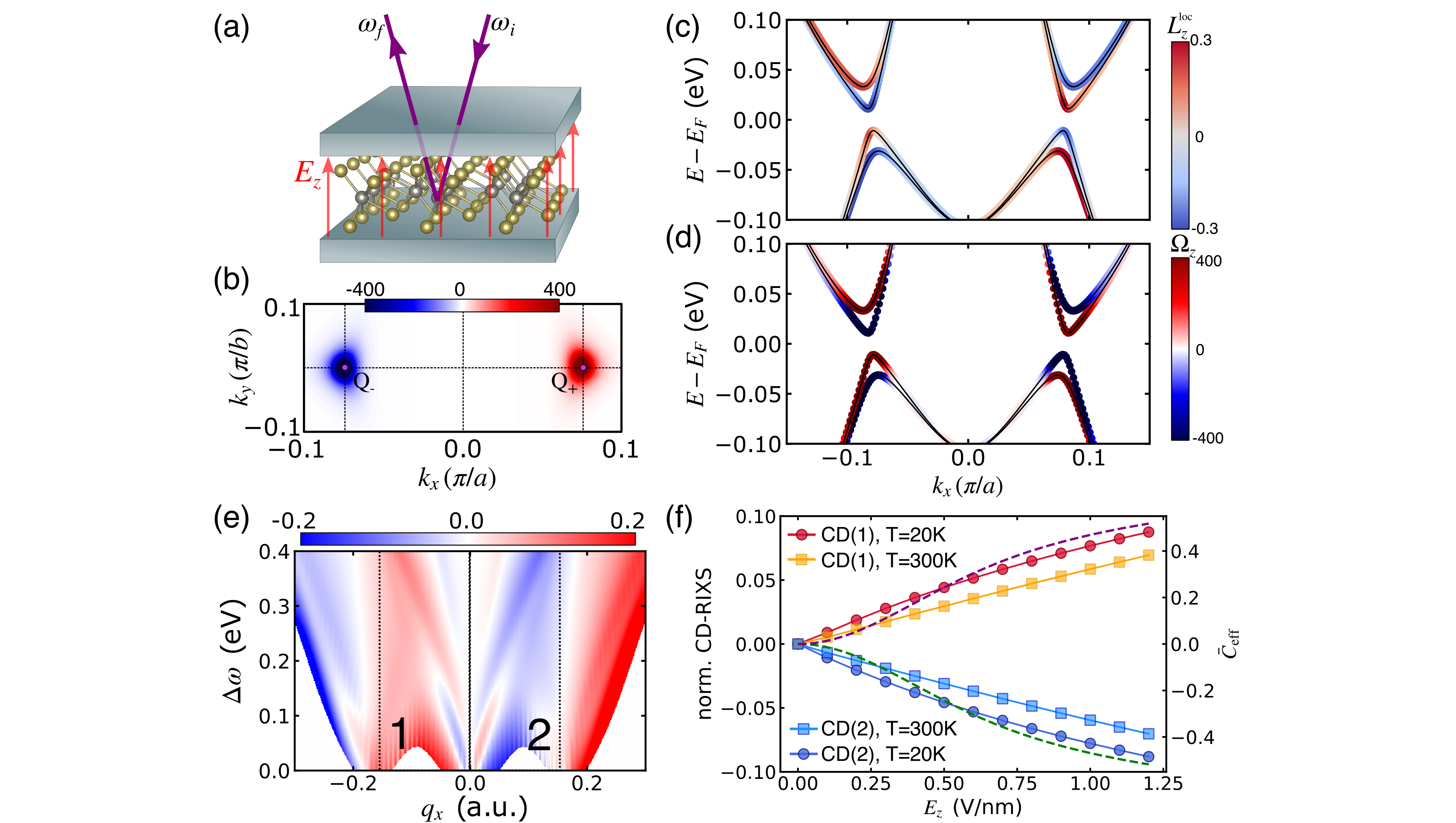}
\caption{\textbf{Field-induced Berry curvature and circular dichroism from 1T$^\prime$-MoS$_2$.} 
(a) Sketch of the setup for observing electrically switchable Berry curvature by RIXS. (b) Berry curvature $\Omega_z$ of the bottom conduction bands in the vicinity of the $Q_\pm$ points for $E_z=0.4$~V/m. (c) Fat-band representation of the the local OAM $L^\mathrm{loc}_z$ (color map) along the X-$\Gamma$-X path for $E_z=0.4$~V/m. (d) Similar to (c), but the color indicates the Berry curvature $\Omega_z$. (e) Normalized CD-RIXS signal $(I_\mathrm{LCP} - I_\mathrm{RCP}) / (I_\mathrm{LCP} + I_\mathrm{RCP} )$ for $2\theta=130^\circ$ and unpolarized outgoing photons. The rectangles denoted by 1 and 2, respectively, represent the phase space where almost-direct transitions to the bottom conduction band (CB) are the dominant processes. (f) Integrated CD-RIXS signal (normalized over the total intensity) in the  regions of interest 1, 2 in (e) (symbols), compared to the Berry curvature $\bar{C}_\mathrm{eff}$ of the bottom CB integrated region of the BZ shown in (b).}
\label{fig:tpmos2_cd}
\end{figure*}

One of the most remarkable properties of 1T$^\prime$-MoS$_2$ is the tunable Berry curvature upon applying an out-of-plane electric field (see Fig.~\ref{fig:tpmos2_cd}(a))  which can be realized in a hetereostructure~\cite{xu_electrically_2018-1}.
With a sizable electric field $E_z$, the inversion symmetry is broken, inducing an imbalance of the otherwise equivalent Mo atoms. As a result the spin degeneracy is lifted. The resulting spin texture is locked to the OAM texture, which is the source of the emerging Berry curvature (Fig.~\ref{fig:tpmos2_cd}(b)). Time-reversal symmetry dictates the total Berry curvature to vanish. The system acquires a Berry curvature dipole which results in a nonlinear Hall response~\cite{sodemann_quantum_2015}. 
As shown in Fig.~\ref{fig:tpmos2_cd}(b), the bottom conduction band (CB) shows strong Berry curvature (which is opposite for the now spin-split bands), which is almost proportional to the OAM of the projections of the Bloch states onto 
the Mo sites (see Fig.~\ref{fig:tpmos2_cd}(c)--(d)). The magnitude of the Berry curvature and the Berry curvature dipole grows upon increasing the field strength up to the point where the CB and valence bands (VB) touch. At this point the system undergoes a topological phase transition accompanied by a gap closing at the critical field strength of $E^\mathrm{crit}_z\approx 0.9$~V/nm~\footnote{The value $E^\mathrm{crit}_z =  0.9$~V/nm is smaller than the one obtained in Ref.~\cite{qian_quantum_2014}. We include the external electric field directly without taking any additional screening into account, which slightly overemphasizes the effects of the electric field.}. While the spin-Chern number drops to zero~\cite{qian_quantum_2014}, the (charge) Berry curvature continues to grow upon increasing $E_z$.

Inspecting the CD signal for momentum transfer $\vec{q}$ along the $x$-axis (Fig.~\ref{fig:tpmos2_cd}(e)) we notice pronounced dichroism in the entire $q_x$-$\Delta \omega$ phase space stemming from now strongly asymmetric spectral lobes. In particular, there is a strong CD for small $|q_x|$ at the fringes of the lobes, which is associated with transitions to the kinks of the CB with maximal Berry curvature (highlighted by the rectangles in Fig.~\ref{fig:tpmos2_cd}(e)). Consistent with the scenario (ii) outlined above, the CD signal predominantly originates from the magnetic states of the unoccupied states, while the magnetic quantum number of the occupied orbitals is less relevant. 

Next we compare the CD signal in regions of interest directly to the Berry curvature. 
Since the two CBs are almost degenerate away from $Q_{\pm}$, rendering their opposite Berry curvature indistinguishable, we computed the effective Berry curvature by averaging over an energy window:
\begin{align}
  \label{eq:cbc_eff}
  \bar{C}^{\pm}_\mathrm{eff} = \sum_\alpha \int d\vec{k}\, \Omega_{z,\alpha}(\vec{k}) w_\alpha(\vec{k}) \ ,
\end{align}
where the integration range is the part of the BZ shown in Fig.~\ref{fig:tpmos2_cd}(b). 
The sum over $\alpha$ includes the two CBs, and $w_\alpha(\vec{k})$ is a Gaussian weight factor around the bottom CB. The definition~\eqref{eq:cbc_eff} is analogous to the valley Chern number~\cite{ezawa_valley-polarized_2012}; the averaging procedure reduces $\bar{C}^{\pm}_\mathrm{eff}$ for small $E_z$ and thus almost degenerate CBs, which reflects the suppression of observable experimental effects for $E_z\rightarrow 0$~\cite{xu_electrically_2018-1}. 
Switching on the electric field, the effective integrated Berry curvature~\eqref{eq:cbc_eff} grows approximately linearly in magnitude 
(Fig.~\ref{fig:tpmos2_cd}(f)); $\bar{C}^{+}_\mathrm{eff}$ ($\bar{C}^{-}_\mathrm{eff}$) is positive (negative). 

Strikingly, the CD-RIXS signal shows the same trend. For a direct comparison we integrated the CD signal in the boxes 1, 2 in Fig.~\ref{fig:tpmos2_cd}(f) and normalized by the corresponding polarization-averaged intensity. At low temperature (circles in Fig.~\ref{fig:tpmos2_cd}(f)), the CD signal is very close to the integrated Berry curvature. The non-monotonic behavior of the total Berry curvature is also captured. Increasing the temperature to $T=300$~K -- which is on the order of the of band gap -- the normalized CD-RIXS signal is only slightly reduced. 

Similar to MoSe$_2$, the finite momentum transfer selects the phase space of the particle-hole excitations. For small $q_x>0$ 
($q_x < 0$), transitions from the VB to the CB are only possible close to $Q_+$ ($Q_-$). For this reason, excitations occur only in regions with positive (negative) Berry curvature, thus providing momentum-resolved topological properties.

\section*{Discussion}

We have presented calculations of CD-RIXS from molybdenum-based TMCDs, in particular monolayer MoSe$_2$ and the QSHI 1T$^\prime$-MoS$_2$. CD-RIXS is sensitive to the OAM of the orbitals involved in the particle-hole excitations in the Mo $d$ manifold; the magnetic character of the conduction band plays the dominant role. Unlike simple magnetic materials, the relevant bands possess an OAM texture, i.\,e. a momentum dependence of the relative contributions of the magnetic $d$ orbitals. The OAM texture is a signature of Berry curvature, which renders CD-RIXS a powerful tool to investigate quantum geometric and topological properties of materials. Even for nonmagnetic materials with vanishing total Berry curvature, the CD signal can be pronounced at finite momentum transfer, which gives CD-RIXS an advantage over optical spectroscopies such as Raman spectroscopy. Furthermore, the orbital character of the unoccupied bands is the predominant factor determining the CD. Hence, CD-RIXS yields insights into the OAM texture of the conduction bands. This is a clear advantage over ARPES, as access to the conduction bands (and their orbital properties) is only possible within pump-probe photoemission~\cite{volckaert_momentum-resolved_2019-1,puppin_excited-state_2022}. Moreover, extracting the OAM from CD-RIXS is straightforward due to the selection rules. In contrast, the manifestion of Berry curvature in CD-ARPES is much more involved, as extrinsic effects such as the final-state effects or the experimental geometry complicate the interpretation.
The insensitivity of the photons to external electric fields allows to study field-induced transitions, as demonstrated by the switchable Berry curvature in 1T$^\prime$-MoS$_2$.
The site specificity of RIXS also provides information on the localization of the Bloch wave-function~\cite{lee_metrology_2021}, which is directly connected to the band topology~\cite{cano_band_2021}.

The idea of selecting the phase space of particle-hole excitations by controlling the momentum transfer $\vec{q}$ is general and can be applied to many other matierals. The only requirement is an anisotropic band structure with respect to the direction of $\vec{q}$. This is generically the case in systems with broken inversion symmetry (which is required for nonvanishing Berry curvature if time-reversal symmetry is present). 
For the Berry curvature to be reflected in the local OAM, localized orbitals are required, as is typically the case for bands with $d$ or $f$ orbital character. Investigating topological properties with CD-RIXS is thus expected to be applicable to a large class of materials.

RIXS typically probes a number of many-body excitations besides the particle-hole transitions from the valence to the conduction bands. The main challenge for extracting the information on the OAM texture will be the separation of the dispersive particle-hole continuum from other excitations. Local excitations are typically reflected in a strong RIXS signal below the band gap; their contribution can be isolated from the particle-hole $d$--$d$ transitions by careful analysis~\cite{paris_strain_2020}. For TMDCs in particular, excitons are pronounced but delocalized in space~\cite{dong_direct_2021}, which should reduce their spectral weight in RIXS spectra.
Furthermore, inelastic scattering and relaxation processes in the valence band result in a delocalized response of the electronic structure termed fluorescence, which can be dominant, especially in metallic systems. Removing the fluorescence background is challenging but possible~\cite{zhou_persistent_2013}; for molybdenum-based compounds it has been demonstrated that the main spin-orbit-split peaks are visible on top of the fluorescence line~\cite{thomas_resonant_2015}, which supports the feasibility of the proposed CD-RIXS experiment. Apart from magnetic particle-hole excitations, potentially studying the chirality of many-body excitations such as excitons~\cite{caruso_chirality_2021} with CD-RIXS is an interesting perspective.

The recent development of time-resolved RIXS~\cite{mitrano_probing_2020,parchenko_orbital_2020,paris_probing_2021} underlines the potential for tracing out-of-equilibrium phenomena. In parallel, several realistic theoretical proposals for using time-resolved RIXS to study light-driven materials by RXIS~\cite{,chen_theory_2019,chen_observing_2020,wang_x-ray_2021} have been put forward. Combining our CD-RIXS analysis with time-resolved RIXS is thus expected to open a new route for exploring light-induced topological phase transitions~\cite{sie_ultrafast_2019,hubener_creating_2017}.

\section*{Acknowledgments}

We thank Urs Staub for fruitful discussions and useful insights.
The calculations have been performed at the Merlin6 cluster at the Paul Scherrer Institute.
M.S. thanks the Swiss National Science Foundation SNSF for its support with an Ambizione grant (project no.~193527). 
T.S. acknowledges support from SNSF grant No.~200021\_207904 and  
P.W. support from SNSF Grant No. 200021-196966.
This research was supported by the NCCR MARVEL, a National Centre of Competence in Research, funded by the Swiss National Science Foundation (grant number 205602).

\section*{Methods}

\subsection*{Calculation of the RIXS cross section\label{app:hartree_rixs}}

We compute the RIXS intensity from the Kramers-Heisenberg formula in the language of many-body states:
\begin{align}
  \label{eq:kramer_heisenberg}
  I(\omega_i,\vec{q}_i,\omega_f,\vec{q}_f) = \sum_f \left|A_{fi}(\omega_i,\vec{q}_i,\vec{q}_f)\right|^2 \delta(E_i + \Delta \omega - E_f) \ .
\end{align}
Here, $E_i$ ($E_f$) denotes the energy of the ground (final) state, while $\Delta\omega = \omega_i - \omega_f$ is the energy transfer. The RIXS amplitude $A_{fi}(\omega_i,\vec{q}_i,\vec{q}_f)$ is defined by
\begin{align}
  \label{eq:rixs_ampl}
  A_{fi}(\omega_i,\vec{q}_i,\vec{q}_f) = \sum_{\vec{R}}e^{-i \vec{q}\cdot \vec{R}} \sum_x \frac{\langle \Psi_f | \hat{\Delta}^\dagger_{f,\vec{R}}|\Psi_x \rangle \langle \Psi_x| \hat{\Delta}_{i,\vec{R}}|\Psi_i\rangle}{E_x -E_i - \omega_i - i \Gamma} \ ,
\end{align}{}
where $x$ labels all intermediate excited states $|\Psi_x\rangle$ with energy $E_x$; the light-matter interaction with respect to the incoming (outgoing) photon at lattice site $\vec{r}$ is described by $\hat{\Delta}_{i,\vec{R}}$ ($\hat{\Delta}_{f,\vec{R}}$). The momentum transfered to the material is denoted by $\vec{q}= \vec{q}_i - \vec{q}_f$.

The many-body states in Eq.~\eqref{eq:rixs_ampl} are calculated from a Hamiltonian composed of band electrons ($\hat{H}_b$), core electrons ($\hat{H}_{c}$), and their interaction ($\hat{H}_\mathrm{int}$):
\begin{align}
  \label{eq:ham_tot}
  \hat{H} = \hat{H}_b + \hat{H}_{c} + \hat{H}_\mathrm{int} \ .
\end{align}
The Hamiltonian $\hat{H}_b$ describing the band electrons is constructed in the relevant orbitals space from density-functional theory (DFT) as detailed below. The core electrons are described by
\begin{align}
  \label{eq:ham_core}
  \hat{H}_c &= \lambda_\mathrm{SOC} \sum_{\vec{R}} \sum_{m m^\prime}\sum_{\sigma \sigma^\prime} \langle \ell_c m \sigma | \hat{\vec{L}}\cdot\hat{\vec{S}} | \ell_c m^\prime\sigma^\prime\rangle d^\dagger_{\vec{R}m\sigma} d_{\vec{R}m^\prime \sigma^\prime} \nonumber \\ &\quad + 
  E_c \sum_{\vec{R}} \sum_{m \sigma}\hat{n}^c_{\vec{R}m\sigma} \ ,
\end{align}
where $d^\dagger_{\vec{R}m\sigma}$ ($d_{\vec{R}m\sigma}$) is the creation (annihilation) operator of a core electron in state with magnetic quantum number $m$ and spin $\sigma$. The energy levels are solely determined by the spin-orbit coupling $\lambda_\mathrm{SOC}$, the angular momentum quantum number $\ell_c$ and the energy shift $E_c$. We adjust $\lambda_\mathrm{SOC}$ and $E_c$ such that the core levels reproduce tabulated values for the edge energies~\cite{clementi_atomic_1963,clementi_atomic_1967}.

The interaction of core and valence electrons is parameterized by
\begin{align}
  \label{eq:ham_int}
  \hat{H}_\mathrm{int} = - U^c \sum_{\vec{R}}\sum_{j}\sum_{m\sigma} n^b_{\vec{R}j}d_{\vec{R}m\sigma} d^\dagger_{\vec{R}m\sigma} \ ,
\end{align}
where the sum over valence orbitals $j$ is restricted to the atoms where the core hole is created; $n^b_{\vec{R}j}$ denotes the density operator of the band electrons.
We treat the interaction~\eqref{eq:ham_int} on the Hartree-Fock (HF) level, which incorporates the basic physics of bound core-valence excitons~\cite{sangalli_ab-initio_2018,perfetto_first-principles_2016,murakami_ultrafast_2020-1}. Consistent with the HF approximation, the initial and the final states are computed without the interaction term: $(\hat{H}_b + \hat{H}_{c} )|\Psi_{i,f}\rangle = E_{i,f} |\Psi_{i,f}\rangle $. Hence $|\Psi_{i,f}\rangle$ is constructed as a determinant of the occupied Kohn-Sham states. The intermediates states can be constructed as a superposition of single-particle excitation from the initial state:
\begin{align}
  \label{eq:intermediate}
  |\Psi_x\rangle = \sum_{\alpha\nu} A^x_{\alpha\nu}(\vec{k})c^\dagger_{\vec{k}\alpha} d_{\vec{p}\nu} \ ,
\end{align}
where $c^\dagger_{\vec{k}\alpha}$ is the fermionic creation operator with respect to the Bloch state of the valence band $(\vec{k}\alpha)$, while $d_{\nu}$ is the annihilation operator with respect to the core states $(\vec{p}\nu)$. Inserting the ansatz~\eqref{eq:intermediate} into the Schr\"odinger equation $\hat{H}|\Psi_x\rangle = E_x |\Psi_x\rangle$ allows to solve for the exciton amplitudes $A^x_{\alpha\nu}(\vec{k})$. 
More details are presented in the Supplementary Materials~\cite{supplement}.

\subsection*{First-principles implementation\label{app:abinitio}}

We performed DFT calculations with the {\sc Quantum Espresso} code~\cite{giannozzi_quantum_2009} at the level of the Perdew-Burke-Ernzerhof (PBE) approximation to the exchange-correlation functional. We used the corresponding full relativistic pseudopotentials from the {\sc PseudoDojo} project~\cite{van_setten_pseudodojo_2018}. The ground state calculations were performed on a $12\times 12$ Monkhorst-Pack grid of the first Brillouin zone using a plane-wave cutoff of $80$~a.u. and a density cutoff of $500$~a.u. in a supercell size of $50$~a.u. in the out-of-plane direction. We constructed projective Wannier functions (PWFs) using the {\sc Wannier90} code~\cite{pizzi_wannier90_2020}, including the Mo-$d$ and the chalcogen $p$ orbitals on a $15\times 15$ Monkhorst-Pack grid.

% The PBE functional does not capture the band gap of 1T$^\prime$-WTe$_2$ correctely, as exchange effects are important~\cite{tang_quantum_2017,marrazzo_relative_2019}. To correct this slight deficiency we include a shift of the conduction bands (scissor operator) in the Wannier Hamiltonian to match the band gap of $\approx 30$~meV. The out-of-plane electric field $E_z$ was implemented by the saw-like potential in a large supercell; we computed the self-consistent electronic structure and the Wannier functions for each value $E_z=0.2,0.4,0.6,0.8,1.0$~V/m.

This procedure yields the Wannier representation used to contruct the band Hamiltonian $\hat{H}_b$. Consistent with the choice of the PWFs we represent the Wannier orbitals as Slater-type wave-functions:
\begin{align}
  \label{eq:pwf_orb}
  \phi^b_{j}(\vec{r}) = R_{n_j}(Z_j;r)X_{\ell_j m_j}(\Omega_{\vec{r}}) \ ,
\end{align}
where $R_{n_j}(Z_j;r)$ is a hydrogenic radial function with principal quantum number $n_j$ and effective charge $Z_j$, while $X_{\ell m}(\Omega_{\vec{r}})$ denotes the real spherical harmonics. 

Similarly, we describe the core states by the atomic orbitals
\begin{align}
  \label{eq:corewf_orb}
  \phi^c_{m}(\vec{r}) = R_{n_c}(Z_c;r) X_{\ell_c m}(\Omega_{\vec{r}}) \ .
\end{align}
The principal quantum number and the effective charge for the molybdenum core electrons are taken from Refs.~\cite{clementi_atomic_1963,clementi_atomic_1967}. With orbitals~\eqref{eq:pwf_orb} and \eqref{eq:corewf_orb} we compute the optical matrix elements
\begin{align}
  \label{eq:mel_expl}
  M_{jm}(\vec{e}_a,\vec{q}_a) = \int d\vec{r}\, e^{-i\vec{q}_a\cdot\vec{r}} \phi^b_{j}(\vec{r})  \vec{e}_a\cdot\hat{\vec{p}} \phi^c_{m}(\vec{r})
\end{align}
by expanding the exponential $e^{-i\vec{q}_a\cdot\vec{r}}$ into spherical harmonics, using the Clebsch-Gordan algebra, and calculating the remaining radial integrals. The light-matter coupling operators entering the RIXS amplitude~\eqref{eq:rixs_ampl} are then expressed as many-body operators by
\begin{align}
  \hat{\Delta}_{a,\vec{R}} = \sum_{j}\sum_{m\sigma} M_{jm}(\vec{e}_a,\vec{q}_a) c^\dagger_{\vec{R}j\sigma}d_{\vec{R}m\sigma} \ ,
\end{align}
where $a=i,f$ and where $c^\dagger_{\vec{R}j\sigma}$ stands for the creation operator of the band electrons.

The electric field $E_z$ was included by adding the dipole term into the Kohn-Sham Hamiltonian:
\begin{align}
  \hat{H}_b(E_z) = \hat{H}_b(E_z = 0) - q E_z \sum_{\vec{k}}\sum_{\alpha\alpha^\prime} D^z_{\alpha\alpha^\prime}(\vec{k}) c^\dagger_{\vec{k}\alpha}c_{\vec{k}\alpha} \ .
\end{align}
Here, $D^z_{\alpha\alpha^\prime}(\vec{k})$ is the dipole matrix element calculated directly from the Wannier functions~\cite{schuler_gauge_2021-1}.
% $\hat{H}_b(E_z)$. 
Compared to a self-consistent calculation which explicitly includes the electric field, this approach neglects screening effects due to the rearrangement of the density. The impact of the electric field is thus stronger than in reality; comparing to the critical field strengths for the topological phase transition from Ref.~\cite{qian_quantum_2014}, we find roughly a factor of 1.5. We stress that upon rescaling the field strength, excellent agreement with the first-principles electronic structure is obtained.
Due to the strong Coulomb potential, the electric field can be neglected for the core electrons.

\section*{Competing Interests}

The the Authors declare no Competing Financial or Non-Financial Interests.

\section*{Data Availability}

The input files for \textsc{Quantum Espresso} and \textsc{Wannier90}, the simulated RIXS spectra, and the scripts to generate the plots in this work are available on the Materials Cloud in the archive \href{https://archive.materialscloud.org/record/2022.149}{DOI:10.24435/materialscloud:ck-7m}. The custom computer code used to calculate the RIXS spectra is available upon reasonable request. Band structures, the Berry curvature and orbital angular momentum were computed using the open-access code \href{https://github.com/michaelschueler/dynamics-w90}{dynamics-w90}.

\section*{Author Contributions}

P.W. and M.S. perceived the project idea. M.S. performed the calculations and drafted in initial version of the paper. T.S. provided guidance on experimental aspects. All authors discussed manuscript and finished it together.

% \bibliographystyle{apsrev4-1}
% \bibliography{mylib,supp}


\end{document}